\documentstyle[12pt]{article}

\def\D{{\rm d}}    
\def\e{{\rm e}}    
\title{QED RADIATIVE CORRECTIONS IN EXCLUSIVE $\rho^0$ LEPTOPRODUCTION}
\author{Krzysztof Kurek}
\date{}
\begin{document}

\maketitle
\noindent
{\it Institute of Physics, Warsaw University Branch, Lipowa 41,
15-424 Bia\l{}ystok, Poland,\\
 e-mail: Kurek@fuw.edu.pl}\\

\medskip\medskip\medskip
\vskip1cm
\begin{abstract}
The semi-analytical approach to the model independent leptonic
QED corrections to exclusive $\rho^0$ meson leptoproduction 
(i.e. electron and muon scattering experiments) is presented. 
The corrections to $\rho^0$ production at large $Q^2$ as well as
to $\rho^0$ photoproduction are studied in details.
The numerical results are calculated for two different
experimental analyses: NMC (muoproduction at large $Q^2$) 
and ZEUS at HERA (photoproduction). It is shown that the corrections 
are 2-5 \% for NMC and below 2 \% for the ZEUS measurement.
The application of the presented approach to other vector mesons
production is straightforward.

\end{abstract}
\medskip \medskip \medskip
\section{Introduction}
\medskip \medskip
Elastic production of $\rho^0$ mesons by photons, $\gamma p \rightarrow 
\rho^0 p$, has been extensively studied in the fixed target experiments 
\cite{expe,nmc,foto} 
and recently also at HERA $ep$ collider 
\cite{zeus,zeus2,h1}, using virtual and real or quasi-real photons.
In real photoproduction and small $Q^2$ electroproduction of $\rho^0$
meson characteristic features of diffractive processes
are observed. It allows to describe the process in the framework of
the Vector Meson Dominance Model (VMD) and pomeron exchange \cite{vdm}. 
At large $Q^2$ also
the perturbative QCD mechanism (exchange of two gluons between nucleon
and the virtual $q \bar{q}$ pair from photon) may be appropriate. 
Therefore the study of the vector mesons production in DIS may help to
understand the structure of the pomeron and its connection with
perturbative QCD.

\medskip
To determine the one photon exchange (Born) cross section 
in DIS from
the measured cross section the radiative corrections procedure is needed
\cite{cobar,my,cobar1}.
The similar procedure should be applied to extract the Born cross section
for $\rho^0$ meson production. The Feynman diagram in Born approximation
for the reaction is shown in Fig. 1.
\medskip
Unfortunately the problem of radiative corrections in the
case on vector mesons production has not been fully solved \cite{nmc,zeus}.
As the cuts used to which define the events selection 
depend on the experimental setup,
the radiative processes should be taken into account in Monte Carlo 
simulations. In this paper we propose a simpler approach which allows
to estimate an order of magnitude of the radiative effects in the case
of vector mesons production in DIS (high $Q^2$~) and in 
photoproduction. The method is based on the existing
calculations of the radiative corrections for standard DIS events
\cite{cobar,my,cobar1} and on the several, well-justified assumptions
about the features of the vector mesons production.
\medskip
The assumptions are the following:
\begin{itemize}
\item The main contribution to the radiative corrections proceeds from
the virtual (closed loops) and real radiations from leptons.
It means that we can neglect the radiation from hadron (lower) part of 
the diagram in Fig. 1.
\item The radiative processes connected with leptons are exactly the
same as in the "standard" DIS events. The 
events with $\rho^0$ are a subsample of the DIS events observed
in the experiments, selected in the off-line analysis by extra cuts
and selections.
\item The $\rho^0$ cross section can be factorized as a model-independent
lepton part (the flux of virtual or quasi-real photons)  and hadron part
dependent on the mechanism of the vector mesons production (different
models of the production).
\item In both kinematical regions (small or large $Q^2$) the
energy-momentum transver, t, from the photon to $\rho^0$ is assumed
to be small (it can be related to $p_T^2$ of $\rho^0$). This
assumption is justified by all diffractive models \cite{vdm} 
\footnote{There are also some models where $t$ is expected to be
relatively large \cite{ryskin}; however large t is not observed in the experiments
considered in this paper.}
and also confirmed experimentally \cite{nmc,zeus,zeus2,h1}.
\end{itemize}

\medskip
The paper is organized as follows. 
In Section 2 the kinematics are defined and cross sections formulae are
 given. Sections 3, 4 and 5, together with appendices A and B  contain
a description of the application of the standard DIS radiative corrections
procedure for the vector mesons production for large $Q^2$ and
for photoproduction (quasi-real photons), respectively.
A complete set of formulae is given for each kinematical regime.
In Section 6 and 7 the numerical results are given 
for the NMC quasi-elastic $\rho^0$ cross section measurement \cite{nmc} and 
for the ZEUS 
$\rho^0$ photoproduction analysis.
Finally a summary is given in Section 8. 

\medskip\medskip\medskip
\section{Basic Definitions and Kinematics of exclusive vector mesons 
production}
As an example of vector meson we will consider $\rho^0$.
The exclusive $\rho^0$ meson leptoproduction reaction is (Fig. 1)
\begin{equation}
l + N \rightarrow l + N + \rho^0 .
\end{equation}
The definitions of kinematical variables are the following:\\

\medskip\medskip
\begin{tabular}{cl}
$k, k'$ &- four-momenta of the incident and scattered \\
& leptons (electrons or muons ) \\
$p$ & - four-momentum of the target nucleon \\
$v$ & - four-momentum of the vector meson ($\rho^0$)\\
$q = k - k'$ & - four-momentum of the virtual photon \\
$Q^2 = - q^2$ & - an invariant mass squared of the virtual photon \\
$\nu =(p\cdot q)/M$ & - an energy of the virtual photon in the laboratory
system \\
& (M is the proton mass)\\
$x = Q^2/(2 M \nu)$ & - Bjorken scaling variable \\
$y = (p\cdot q)/(p\cdot k)$ & - fraction of the lepton energy lost 
in laboratory \\
& system \\
$W^2 = (p + q)^2$ & - total energy squared in the $\gamma^{\star} N$
system \\
$t = (q-v)^2$ & - four-momentum transfer between the virtual \\
& photon and the vector meson ($\rho^0$)\\
$M_X^2 = (p + q - v)^2$ & - missing mass squared of the undetected
recoiling system\\
$I = (M_X^2 - M^2)/W^2$ & - Inelasticity \\
\end{tabular}

\medskip\medskip
It is also convenient to introduce $S$ variable defined as
$S=2 (p\cdot k)$. Hence we have $Q^2 = x y S$ .

\medskip
Inelasticity $I$ is equal to 0 for exclusive $\rho^0$ production and 
the cuts on inelasticity were used 
to select exclusive $\rho^0$ sample in the NMC analysis \cite{nmc}.
\medskip

The cross section for DIS  in Born approximation can be expressed
in the following way
\begin{eqnarray}
\sigma^B (y,Q^2) \equiv{\D^2\sigma^B (lp \rightarrow lX )\over \D Q^2\D y}=
\Gamma_T^B  \sigma^B_{\gamma^\star} (y,Q^2),
\label{basic}
\end{eqnarray}
where
\begin{eqnarray}
\Gamma_{T}^{B} = \Gamma_{T}^{B} ({\cal S}_{1}^{B},{\cal S}_{2}^{B}) =
\frac{\alpha}{2 \pi Q^2} \frac{y}{x} (1 - x) \left ({{{\cal S}_{1}^B} \over {S x}} +{{{\cal S}_{2}^B}
\over {S (y S+4 M^2 x)}}\right )
\label{flux}
\end{eqnarray}
and
\begin{eqnarray}
\sigma^B_{\gamma^\star} (y,Q^2) \equiv {\D^2\sigma^B (\gamma^\star p \rightarrow X )\over \D Q^2\D y} = \sigma_T + \varepsilon^B \sigma_L.
\label{sig} 
\end{eqnarray}

\medskip
Introducing the function $R = \sigma_L$/$\sigma_T$ 
 the formula
(\ref{sig}) can be rewritten as follows: 
\begin{eqnarray}
\sigma^B_{\gamma^\star} (y,Q^2) =  \sigma_T ( 1 + \varepsilon^B R) ,
\label{sig1}
\end{eqnarray}
where
\begin{eqnarray}
\frac{1}{\varepsilon^B} &=& 1 + \frac{{\cal S}_{1}^{B}}{{\cal S}_{2}^{B}}
(4 M^2 +\frac{y}{ x} S).
\label{epsi}
\end{eqnarray}
The functions ${\cal S}_{1}^B$ and ${\cal S}_{2}^B$ are defined as:
\begin{eqnarray}
{\cal S}_{1}^B (y,Q^2) &=& Q^2 - 2 m_l^2 \nonumber \\
{\cal S}_{2}^B (y,Q^2) &=& 2 \left [(1-y) S^2 - M^2 Q^2 \right ].
\label{esy}
\end{eqnarray}

\medskip
Here $\sigma_T$ and $\sigma_L$ denote the total cross sections for 
photoabsorption of the longitudinaly
and transversaly polarized virtual photon respectively, and can be 
expressed in terms of nucleon (proton) electromagnetic structure
functions
\begin{eqnarray}
\sigma_T &=& {{4 \pi^2 \alpha }\over {Q^2}}  2 x F_1 (x,Q^2), \nonumber \\
\sigma_L &=& {{ 4 \pi^2 \alpha }\over {Q^2}} \left[ (1 + {{4 M^2 x^2}\over {Q^2}})  F_2 (x,Q^2) - 2 x F_1 (x,Q^2) \right ], \nonumber \\
R &=& {{\sigma_L }\over {\sigma_T}} = {{1+{{4 M^2 x^2}\over {Q^2}}}\over {2 x}} {{F_2}\over {F_1}} -1 .
\label{ery}
\end{eqnarray}

\medskip
The similar set of definitions of the cross sections for $\rho^0$ meson production 
can be introduced. We have respectively

\begin{eqnarray}
\sigma^B_{\rho^0} (y,Q^2) \equiv {{\D^2\sigma^B (lp \rightarrow lp\rho^0 )}\over{ \D Q^2\D y}} =
\Gamma_T^B  \sigma^B_{\gamma^\star \rho^0} (y,Q^2).
\label{basicrho}
\end{eqnarray}

\medskip
The $\sigma^B_{\gamma^\star \rho^0}$ is a cross section for the exclusive  $\rho^0$ meson production in virtual photon - nucleon reaction.

\medskip
\begin{eqnarray}
\sigma^B_{\gamma^\star \rho^0} (y,Q^2) \equiv {{\D^2\sigma^B (\gamma^\star p 
\rightarrow p \rho^0 )}\over {\D Q^2\D y}}= 
\sigma_T^{\rho^0} (1 + \varepsilon^B R_{\rho^0}) .
\label{sigrho} 
\end{eqnarray}

\medskip
The cross sections given above corespond to the Born approximation ( ie.
are one-gamma exchange approximation). The measured cross sections
contain also the contributions from the radiative processes and therefore
differ from the Born ones. The measured cross sections 
will be written without {\it B} superscript
\begin{eqnarray}
\sigma (y,Q^2) &\equiv& {\D^2\sigma_{\rm meas} (lp \rightarrow lX )\over \D Q^2\D y}, \nonumber \\
\sigma_{\rho^0} (y,Q^2) &\equiv& {\D^2\sigma_{\rm meas} (lp \rightarrow lp\rho^0 )\over \D Q^2\D y}.
\label{cross}
\end{eqnarray}

\medskip\medskip\medskip
\section{ Radiative corections for exclusive production  
 of vector mesons in deep inelastic scattering.}

\medskip

To estimate the effect of radiative processes we apply the so-called
Dubna radiative corrections scheme, (D scheme), calculated by A.A. Akhundov et al.\cite{cobar,cobar1}.
The detailed description of the calculations can be found
in the original paper (\cite{cobar}) and also in review articles
\cite{cobar1,my}.
The notation used here is based on \cite{my}. 

\medskip
The measured cross section for DIS is now expressed as follows
\begin{eqnarray}
\sigma (y,Q^2) = \sigma^B (y,Q^2)\left(\e^{-\delta_R(y,Q^2)} +
 \delta^{VR}(y,Q^2) \right)
+ \sigma_{\rm in. tail}(y,Q^2) - \sigma^{IR}(y,Q^2) .
\label{dubna}
\end{eqnarray}
\medskip
The $\delta _R$ in eq.~(\ref{dubna}) is responsible for those parts of the 
{\it soft} and {\it hard collinear} photon emissions which could
be resummed to all orders using the covariant exponentiation procedure,
\cite{cobar,cobar1,my}. It reads
\begin{equation}
\delta_{R}=-{\alpha\over{\pi}}\left(\ln {Q^2 \over m^2}-1\right)
\ln{y^2(1-x)^2 \over (1-yx)(1-y(1-x))},
\label{delinf}
\end{equation}
where $m$ is the lepton mass.
\noindent
The $\delta^{VR}$ correction factor in eq.(\ref{dubna}) is a remnant
of the exponentiation and of the subtraction procedure used to disentangle
the infrared divergent terms from the $\sigma_{\rm in. tail}$ cross section.
It thus contains
the vertex correction and is given by
\begin{equation}
\delta^{VR}=\delta_{\rm vtx}-{\alpha\over{2 \pi}}
\ln^2{(1-yx) \over (1-y(1-x))} +
\Phi\left[{(1-y) \over (1-yx)(1-y(1-x))} \right] - \Phi(1),
\label{delrem}
\end{equation}
where  $\Phi (x)$ denote the Spence function and
$\delta_{\rm vtx}$ is given by the formula
\begin{eqnarray}
\delta_{\rm vtx}&=&{2\alpha\over \pi}\left(-1
+ {3\over 4}\ln{Q^2\over m^2}\right) .
\label{mtvtx}
\end{eqnarray}

\medskip
The $\sigma_{\rm in. tail}$ in eq.~(\ref{dubna}) describes the
inelastic radiative tail for the {\it lepton current correction} only

\begin{eqnarray}
\sigma_{\rm in. tail}(y,Q^2) \equiv {\D^2\sigma_{\rm in. tail} \over \D Q^2\D y} &=& {{2\alpha^3 } \over {S }} 
\int \!\!\int \D Q^2_h  \D M^2_h
{{1} \over {Q^4_h}} \Biggl\{ 2 F_1 (x_h,Q_h^2) 
{\cal S}_{1}(y,Q^2,y_h,Q^2_h) \nonumber \\
&+&  F_2 (x_h,Q_h^2) {{1} \over{y S}}
{\cal S}_{2}(y,Q^2,y_h,Q^2_h)\Biggr\}.
\label{dtails}
\end{eqnarray}

In this formula $M_h$ is the invariant mass of the final
hadronic system  and the variables bearing a subscript
 '$h$' refer to virtual photon--target
vertex in contrast to the variables measured in the inclusive
electroproduction experiment. For their definitions see Appendix~A where also
the explicit expressions for the radiator functions ${\cal S}_{i}$ are given.

For our purpose it is more convienient to rewrite formula for
$\sigma_{\rm in. tail}$ in the similar way as for $\sigma^B $ in section 2
(eq. (\ref{basic})):

\begin{eqnarray}
\sigma_{\rm in. tail}(x,Q^2) = {{\alpha} \over {\pi}} 
    \int \!\!\int \D Q^2_h  \D M^2_h
\Gamma_T'  \sigma_{\gamma^\star}' .
\label{dtailsmoj}
\end{eqnarray}
In order to simplify  the notation, the "prim" denotes that $\Gamma_T$ and
$\sigma_{\gamma^\star}$ depend not only on $Q^2$,$y$  but also on
$Q^2_h$ and $y_h$  \\
\noindent 
$\Gamma_T'$ and $\sigma_{\gamma^\star}'$ have the forms as in the equations (\ref{flux}) ,(\ref{sig1}) and (\ref{epsi})
but now with the different radiator functions ${\cal S}_{1}$ and ${\cal S}_{2}$ (see Appendix A and \cite{my,cobar1}):
\begin{eqnarray}
\Gamma_T = \Gamma_T ({\cal S}_{1},{\cal S}_{2}) \nonumber \\
\sigma_{\gamma^\star} = \sigma_T (1 + \varepsilon R) 
\end{eqnarray}
and finally
\begin{eqnarray}
\varepsilon = \varepsilon ({\cal S}_{1},{\cal S}_{2}) 
\label{sisi}
\end{eqnarray}

\medskip
It is well known that the
$\sigma_{\rm in. tail}$ cross section
is infrared divergent. To regularize it a simple trick ('fixation
procedure') \cite{cobar,cobar1} is used.
It consists in adding to $\sigma_{\rm in. tail}$  and subtracting
 from it  an extra term
\begin{equation}
\sigma^{IR}(x,Q^2) = {\D^2\sigma^{IR}\over \D Q^2\D x}
= \sigma^{B} F^{IR} ,
\label{dir}
\end{equation}
where
\begin{equation}
F^{IR} = \int\!\! \int \D Q^2_h \D M^2_h
    {\cal F}^{IR}(y,Q^2,y_h,Q^2_h).
\label{dir1}
\end{equation}
 In the added term
an integration over a full photon phase space was carried out,
resulting in the above given expressions for $\delta _R$ and $\delta ^{VR}$.
The subtracted term appears explicitly in eq. (\ref{dubna})
so that the difference $\sigma_{\rm in. tail} - \sigma^{IR}$  is finite
over the full
kinematic domain of $Q^2_h$ and $M_h^2$.  The function ${\cal F}^{IR}$ is given in Appendix A.

\medskip
The magnitude of the radiative effects in the measured cross
section will be characterized by the so called radiative correction
factor, $\eta (x,y)$, defined as follows
\begin{equation}
\eta (x,y) = {\sigma^B \over \sigma_{\rm meas}}\,.
\label{eta}
\end{equation}
Combining formulae in eqs. (\ref{dubna}), (\ref{dtails}) and (\ref{dir}) the factor
$\eta$ can be expressed as follows
\begin{eqnarray}
{{1}\over {\eta}} = {\e^{-\delta_R} + \delta^{VR} + {{\sigma_{\rm in. tail}}\over {\sigma^B}} - F^{IR}} .
\label{etafinal}
\end{eqnarray}

To simplify the notation the $y$ and $Q^2$ dependence in eq. (\ref{etafinal})
is not shown explicitely.
\medskip
Finally the vacuum polarisation was taken into
account via the 'running' $\alpha(Q^2)$ which in the $Q^2 \gg {m_f}^2$
approximation ($m_f$ stands for the lepton and quark  masses)
is 
\begin{equation}
 \alpha(Q^2)=\frac{\alpha}{1+  \sum_{f}c_f Q_f^2
\delta_{\rm vac}^f} ,
\label{alfa}
\end{equation}
where $c_{f}$ and $Q_{f}$ are the colour factor and the electric charge of
fermions $f~ (c_{f} = Q_{f} = 1$ for leptons); '$f$' runs over all leptons
and quarks.

\noindent 
$\delta_{\rm vac}^{f}$ in the formula (\ref{alfa}) is equal to
\begin{equation}
\delta_{\rm vac}^{f} = \frac{2 \alpha}{\pi}\left(- \frac{5}{9}
+ \frac{1}{3}\ln{\frac{Q^2}{m_{f}^{2}}}\right) .
\label{vacum}
\end{equation}
As it was said above this formula holds for $Q^2\gg m_{f}^2$. 
The full formula is discussed in the case
of photoproduction (section 4) and is given in Appendix B.

\medskip
Now we can introduce in the same way the $\eta$ factor for exclusive 
vector meson production. Again as an example 
$\rho^0$ meson is considered.
 The $\eta_{\rho^0}$  radiative correction factor is given
in the formula similar to eq. (\ref{etafinal})
\begin{eqnarray}
{{1}\over {\eta_{\rho^0}}} = {\e^{-\delta_R} + \delta^{VR} + {{\sigma^{\rho^0}_{\rm in. tail}} \over {\sigma^B_{\rho^0}}} - F^{IR}}
\label{etarho}
\end{eqnarray}
The $\delta_R$ and $\delta^{VR}$  are calculated exactly in the same 
way as for inclusive  DIS. 
The important differences are connected with the real-photons radiation part (inelastic tail) which is described by
different formula. 
It is not very suprising because this term depends on the structure functions
(or cross sections) and therefore  on the process measured in the experiment.
\medskip
The limits of the integrations in eqs. (\ref{dtails}) and (\ref{dir}) 
 depend on the selection criteria used in the analysis to 
extract the exclusive $\rho^0$ sample.
 Therefore the last term in eq. (\ref{etarho}), $F^{IR}$,
is also
affected, though the formula for the integrand function 
${\cal F}^{IR}$ is universal. 
We will come back to this question in the section 6.
\medskip
Let us consider now the inelastic tail term in formula (\ref{etarho}).
Taking into account eqs. (\ref{basicrho}) and (\ref{dtails}) the ratio
${\sigma^{\rho^0}_{\rm in. tail}}$/${\sigma^B_{\rho^0}}$ can be written 
as follows
\begin{eqnarray}
{{\sigma^{\rho^0}_{\rm in. tail}}\over {\sigma^B_{\rho^0}}} = {{\alpha} \over {\pi}} 
    \int \!\!\int \D Q^2_h  \D M^2_h
{{{\Gamma_T}'  {\sigma^{\rho^0}_{\gamma^\star}}'}\over { \Gamma^B_T \sigma^{B \rho^0}_{\gamma^\star}}} .
\label{ratio}
\end{eqnarray}
The "prim" again means the hidden dependence on "h" 's variables.
According eq. (\ref{sigrho}) $\sigma_T^{\rho^0}$ is equal to
\begin{eqnarray}
\sigma_T^{\rho^0} = 
{{\sigma^{B}_{\gamma^\star \rho^0}} \over {(1 + \varepsilon^B R_{\rho^0})}} 
\label{sigrho1}
\end{eqnarray}
and therefore ${\sigma^{\rho^0}_{\gamma^\star}}'$ can be rewritten as 
\begin{eqnarray}
{\sigma^{\rho^0}_{\gamma^\star}}' = {\sigma^B_{\gamma^\star \rho^0}}'{{(1 + 
{\varepsilon}' R_{\rho^0})} \over {(1 + {\varepsilon^B}' R_{\rho^0})}} .
\label{sigrho2}
\end{eqnarray}
\medskip
Putting above formula to eq. (\ref{ratio}) and then to formula (\ref{etarho})
we obtain finally our "master" formula for $\eta_{\rho^0}$ radiative 
correction factor in the large $Q^2$ regim:
\begin{eqnarray}
{{1}\over {\eta_{\rho^0}}} = \e^{-\delta_R} +\delta^{VR} + {{\alpha} \over {\pi}} \int \!\!\int \D Q^2_h \D M^2_h {{\Gamma_T'} \over {\Gamma_T^B}}  
{{(1 + {\varepsilon}' R_{\rho^0}) }\over {(1 + {\varepsilon^{B}}' R_{\rho^0})}} 
{{{\sigma^{B}_{\gamma^\star \rho^0}}'}\over {\sigma^B_{\gamma^\star \rho^0}}} .
\label{master1}
\end{eqnarray}
\medskip
 As in the previous formulae the hidden notation is used; the $y$  and
$Q^2$ dependence is not shown explicitely as well as ' denotes the 
"h" variables dependence over which the integrations is performed.
\medskip
The photon fluxes ($\Gamma_T'$ and $\Gamma_T^B$ ) and  $\varepsilon 's$
(${\varepsilon^B}'$ and ${\varepsilon}'$) are expressed in terms of 
radiator functions ${\cal S}_{1}^B,{\cal S}_{2}^B,{\cal S}_{1},{\cal S}_{2}$ 
(see Appendix A.). $R_{\rho^0}$ and the cross section 
$\sigma^B_{\gamma^\star \rho^0}$ were measured in the experiment for 
different $y$ and $Q^2$ and in the numerical calculations
the fit to data was used (see \cite{nmc}).
The numerical calculations for NMC analysis are discussed in the section 6.

\medskip\medskip\medskip
\section{ Radiative corections for exclusive vector meson 
photoproduction  }
\medskip
In this section we present the application of the radiative corrections
procedure to the small $Q^2$ region. As an example ZEUS $\rho^0$ 
photoproduction at HERA will be considered (section 7) , where the considered
events have very small $Q^2$, nearly 0 ($\approx 10^{-9}$ \cite{zeus}).

\medskip
The starting point is the cross section formula (\ref{basicrho}). 
Due to the fact that photon emitted by electron is nearly real,
the electron is practically not distorted and therefore not observed
in the experiment (escapes through the pipe). Also proton is not observed in this case. Therefore
to estimate the $\rho^0$ cross section the theoretical input about
production mechanism is needed. In the ZEUS analysis the Vector Meson Dominance model (VMD) relations about transverse and longitudinal cross sections in the limit of $Q^2$ close to 0 were assumed (\cite{zeus,vdm})
\begin{eqnarray}
\sigma_{L \rho^0} (W,Q^2) = {{Q^2}\over {m_{\rho^0}^2}} \sigma_{T \rho^0} (W,Q^2) ,\nonumber \\
\sigma_{T \rho^0} (W,Q^2) = {{\sigma_{\gamma \rho^0} (W)} \over { \left( 1+{{Q^2}\over {m_{\rho^0}^2}} \right ) ^2}} ,
\label{vdmpred}
\end{eqnarray}
where $\sigma_{\gamma \rho^0} (W)$
is the cross section for elastic photoproduction ($Q^2 = 0$) of $\rho^0$
meson. Substituting the VMD predictions (eq. (\ref{vdmpred})) to eq. (\ref{sigrho})
we obtain the following expresion for the $\sigma^B_{\gamma^\star \rho^0}$:
\begin{eqnarray}
\sigma^{B}_{\gamma^{\star} \rho^0} = \sigma_{T \rho^0} (W,Q^2) 
\left( 1+ \varepsilon^B \frac{Q^2 }{m_{\rho^0}^2} \right) = 
\sigma_{\gamma \rho^0} (W) {{\left( 1 + \varepsilon^B \frac{Q^2}{m_{\rho^0}^2} 
\right) }\over {
\left( 1+\frac{Q^2}{m_{\rho^0}^{2}} \right )^2}} .
\label{sigvdm}
\end{eqnarray}
Finally combining above expresion together with eq. (\ref{sigrho}) we 
find that the differential  cross section in Born approximation for $\rho^0$
production is expressed as follows
\begin{eqnarray}
\sigma^B_{\rho^0} (y,Q^2) = \Gamma_T^B \sigma_{\gamma \rho^0} (W) 
{{\left( 1 + \varepsilon^B \frac{Q^2}{m_{\rho^0}^2} \right)} 
\over {\left( 1+\frac{Q^2}{m_{\rho^0}^2} \right )^2}} = \Phi^B (y,Q^2) 
\sigma_{\gamma \rho^0} (W) .
\label{sigfoto}
\end{eqnarray} 
$\Phi$ is so-called  effective photon flux. It reads in Born approximation:
\begin{eqnarray}
\Phi^{B} (y,Q^2) &\equiv& \Phi ({\cal S}_{1}^B,{\cal S}_{2}^B) = \Gamma_T^B 
{{\left( 1 + \varepsilon^B \frac{Q^2}{m_{\rho^0}^2} \right) }\over 
{\left( 1+\frac{Q^2}{m_{\rho^0}^2} \right )^2}} \nonumber \\
 &=& \frac{\alpha}{2 \pi Q^2}  \Biggl[ 
\frac{1+(1-y)^2}{y} -\frac{2 (1-y)}{y} \left ( \frac{Q^2_{min}}{Q^2} - 
\frac{Q^2}{m_{\rho^0}^2}\right ) \nonumber \\
&+& \epsilon \left( \frac{y}{2} \left (1- \frac{Q^2}{m_{\rho^0}^2}\right) -
2 (1-y) \frac{Q^2_{min}}{Q^2} \right) \Biggr] \frac{1}{1+\epsilon} 
\frac{1}{\left( 1+\frac{Q^2}{m_{\rho^0}^2} \right ) ^2} .
\label{fluxfull}
\end{eqnarray}

In this equation $Q^2_{min}$ is the minimal value of $Q^2$ allowed
by kinematics ($10^{-9} GeV^2$ in ZEUS), $m_{\rho^0}$ and $m_e$ are mass of $\rho^0$ meson and electron, respectively.
The $\epsilon$ parameter is equal to $4 M^2 Q^2\over{S^2 y^2}$ and is smaller
than $10^{-6}$ in ZEUS photoproduction events selection and can be neclected.
This simplify the effective photon flux formula to the formula used in
ZEUS paper \cite{zeus}
\begin{eqnarray}
\Phi^B (y,Q^2) = {{\alpha} \over{2 \pi Q^2}} \Biggl[ {{1+(1-y)^2}\over {y}} -
{{2 (1-y)}\over {y}} \left ( {{Q^2_{min}} \over {Q^2}} - {{Q^2} \over {m_{\rho^0}^2}}
\right )  \Biggr]{{1} \over{ 
\left( 1+{{Q^2} \over {m_{\rho^0}^2}} \right ) ^2}} .
\label{fluxhera}
\end{eqnarray}
\medskip
The photoproduction cross section $\sigma_{\gamma \rho^0} (W)$
at $Q^2 = 0$ is obtained as a ratio of the corresponding ep cross section
integrated over $y$ and $Q^2$ ranges covered by the ZEUS measurement and of the effective photon flux integrated over the kinematical range:
\begin{eqnarray}
\sigma_{\gamma \rho^0} = {{\sigma_{\rho^0}} \over {\Phi_{int}}}.
\end{eqnarray}
$\sigma_{\rho^0}$ is determined from data as follows:
\begin{eqnarray}
\sigma_{\rho^0} &=& {{N_{\pi^+\pi^-}} \over{L A}} ,\nonumber \\
\Phi_{int} &=& \int \!\!\int \D Q^2 \D y \Phi^B (Q^2,y) . 
\label{pions}
\end{eqnarray}
$N_{\pi^+\pi^-}$ is the number of observed events with pions from $\rho^0$,
 A and L are
acceptance and luminosity, respectively, see \cite{zeus}.
\medskip
This procedure determines averaged cross section for  elastic $\rho^0$ meson
photoproduction over a given W range.\footnote{The integration over $Q^2$ and
$y$ is experimental analysis dependent; if it is possible to measure
$Q^2$ and $y$ dependent cross section for small $Q^2$ (or $Q^2 \approx 0$) these integrations in the formulae
presented in this section are not needed. }
\medskip
To calculate radiative correction factor $\eta$ for the measured average 
cross section $\sigma_{\gamma \rho^0}$ the same method as for
the large $Q^2$ $\rho^0$ meson production can be used, however now in addition
the integration over $W$ (or $y$) and $Q^2$ ranges where the events are
accepted and contributed to the averaged cross section, must be performed.
As a result we obtain only one number for $\eta$ ( or equivalently "new",
corrected $\sigma_{\gamma \rho^0}$) integrated and averaged
over measured kinematical region. 
\medskip\medskip 
The measured cross section $\sigma_{\rho^0}$ contains not only pions pair from Born $\rho^0$ production process (Fig. 1) but also from
radiative events. Therefore the measured cross section $\sigma_{\rho^0}$ can be written as
\begin{eqnarray}
\sigma_{\rho^0} = \sigma_{\rho^0}^B + \sigma_{\rho^0}^R ,
\label{new}
\end{eqnarray}
where $\sigma_{\rho^0}^B$ is an integrated Born cross section from eq. (\ref{sigfoto}) and $\sigma_{\rho^0}^R$ is the radiative part of the measured
cross section also integrated over $W$ ($y$) and $Q^2$.
\medskip
Therefore the corrected formula for $\sigma_{\gamma \rho^0}$ 
can be written as
\begin{eqnarray}
\sigma_{\gamma \rho^0}^{\rm corr} = {{\sigma_{\rho^0}^B} \over {\Phi_{int}}} = {{\sigma_{\rho^0} - \sigma_{\rho^0}^R }\over {\Phi_{int}}} ,
\label{newrho}
\end{eqnarray}
where the radiative part of the cross section $\sigma_{\rho^0}^R $ is given 
by the similar formula as in eq. (\ref{dubna}):
\begin{eqnarray}
\sigma_{\rho^0}^R  = \int \!\!\int \D Q^2 \D y \left ({\sigma_{\rho^0}^B} \left (
\delta^{VR} -\delta_R +\delta_{\rm anom} \right) + \sigma_{\rm in. tail} - \sigma^{IR}\right) . 
\label{rhorad}
\end{eqnarray}
\medskip
$\sigma_{\rm in. tail}$ is now expressed in terms of effective flux (now {\it
not} in Born approximation) and $\sigma_{\gamma \rho^0}$
\begin{eqnarray}
\sigma_{\rm in. tail} (y,Q^2) = {{\alpha }\over{\pi}} \int \!\!\int \D Q^2_h  \D M^2_h 
\Phi' \sigma_{\gamma \rho^0} (M_h) ,
\label{tailfoto}
\end{eqnarray}
where the effective flux now depends on "h" indexed variables and full
radiative functions ${\cal S}_{1}$ and ${\cal S}_{2}$
\begin{eqnarray}
\Phi' = \Phi'({\cal S}_{1},{\cal S}_{2}) .
\end{eqnarray}
The $\delta^{VR},\delta_R$ and $\sigma^{IR}$ have the same meaning as in 
eq. (\ref{dubna}) and in section 2, but the formulae for these quantities 
are slightly different due to the
fact that we consider now very small $Q^2$ and all quantities
must be calculated exactly keeping all terms, including that with electron's mass.
In the formulae in section 2. the lepton mass terms (like $m_l^2\over{Q^2}$)
were neglected. \\
The correct formulae with electron's mass terms for $\delta^{VR}, \delta_R$
and for vertex and vacuum polarization corrections are collected in Appendix B.
For detailed calculations the reader is referred to
the original papers \cite{cobar,cobar1}. \\
The new contribution, $\delta_{\rm anom}$ , is present in eq. (\ref{rhorad}).
This correction is important only for real (or quasi-real) photons (photoproduction)
and takes into account the contribution from anomalous magnetic moment of electron.
The calculations and the formula for $\delta_{anom}$ are discussed in the
next section in more details.

\medskip
Now it is possible to formulate the expresion for the corrected cross section
for elastic $\rho^0$ meson photoproduction
\begin{eqnarray}
\sigma_{\gamma \rho^0}^{\rm corr} &=& \sigma_{\gamma \rho^0} \Biggl\{ 1 -
\int \!\!\int \D Q^2  \D y \Biggl [ {{\Phi^B \sigma_{\gamma \rho^0} (W)
(\delta^{VR} - \delta_{R} + \delta_{\rm anom})} \over {\sigma_{\rho^0}}}  \nonumber \\
&+& \frac{\alpha}{\pi} \int \!\!\int \D Q^2_h  \D M^2_h  
\frac{\Phi'\sigma_{\gamma \rho^0} (M_h)}{\sigma_{\rho^0}}
- {\cal F}^{IR}\frac{\Phi^B \sigma_{\gamma \rho^0} (W)}{\sigma_{\rho^0}}
\Biggr]\Biggr\} .
\label{nearmaster}
\end{eqnarray}
This equation leads  to our second "master" formula for radiative correction
factor $\eta$, now for the photoproduction limit
\begin{eqnarray}
\eta = {{\sigma_{\gamma \rho^0}^{\rm corr}} \over{ \sigma_{\gamma \rho^0}}}
&=& 1 - \int \!\!\int \D Q^2  \D y \Biggl [ {{\Phi^B \sigma_{\gamma \rho^0}(W)} \over {\sigma_{\rho^0}}} 
(\delta^{VR} - \delta_{R} +\delta_{\rm anom}) \nonumber \\
&+& \frac{\alpha}{\pi}
\int \!\!\int \D Q^2_h  \D M^2_h {{\Phi'\sigma_{\gamma \rho^0} (M_h)} 
\over{\Phi^B \sigma_{\gamma \rho^0}(W)}} - {\cal F}^{IR}\Biggr] .
\label{master2}
\end{eqnarray}
The cross section $\sigma_{\gamma \rho^0}(W))$ ( and 
$\sigma_{\gamma \rho^0}(M_h))$) on the right hand side of the eq. 
(\ref{master2}) results from a fit to the existing data; in the present paper the same
parametrization was used as in the ZEUS analysis \cite{zeus,DL}. 
$\sigma_{\rho^0}$ is the averaged cross section  measured in the experiment.
\medskip
The limits of integrations and the experimental cuts used in our numerical calculations
will be discused in the section 7.\\
The anomalous electron's magnetic moment contribution is presented in the next
section.
  
\medskip \medskip \medskip
\section{The anomalous magnetic moment's radiative correction $\delta_{anom}$}
In this section we consider in more details the effect 
proceeding from anomalous 
magnetic moment of the electron. The cross section contribution from the anomalous magnetic 
moment of electron in the case of DIS is (see e.g. \cite{cobar1})

\begin{eqnarray}
{\D^2\sigma_{\rm anom}\over \D y \D Q^2} &=& {{2 \alpha^{3} m_{e}^{2} \nu_{anom}}\over {
S Q^{6} x}} \left [ 2 \beta^{2} y^2 x F_{1} (x, Q^2) - (2 - y)^{2} F_{2} (x, Q^2) \right] ,
\label{anomal}
\end{eqnarray}

where
\begin{eqnarray}
\nu_{\rm anom} &=&-{{1}\over{\beta}} \ln{{\beta +1}\over {\beta -1}} ,\nonumber \\
\beta &=& \sqrt{1 + {{4 m_{e}^2}\over {Q^2}}} . \nonumber \\
\end{eqnarray}
The anomalous magnetic moment's contribution to the cross section is of order of $1/Q^6$
in contrast to the Born and other radiative cross sections which are of order $1/Q^4$.
It means that this effect is important only for very small $Q^2$, where the ratio
$m_{e}^2 /Q^2$ is large. It is exactly the photoproduction case. \\
It is easy to show that the formula 
(\ref{anomal}) can be rewritten in the following form: 
\begin{eqnarray}
{\D^2\sigma_{\rm anom}\over \D y \D Q^2} &=& {\delta_{\rm anom} 
(1 - \varepsilon_{R}) \sigma^B (y,Q^2)},
\label{anom}
\end{eqnarray}
where $\varepsilon_{R}$ and $\delta_{\rm anom}$ are given by
\begin{eqnarray}
\varepsilon_{R} &=& R {{(\varepsilon^{B} - \varepsilon)}\over 
{1 + \varepsilon^{B} R}} ,\nonumber \\
\delta_{\rm anom} &=& {{\alpha^{2} m_{e}^{2} y^2} \over{2 \pi Q^{6} 
\Gamma_{T}^{B}}}\nu_{\rm anom} \left [y \beta^2 - {{(2 -y)^{2} S}\over {4 M^2 x +y S}} 
\right ] 
\label{deltanom}
\end{eqnarray}
and $\varepsilon^{B}, \Gamma_{T}^{B}$ and $R$ are given by (\ref{epsi}),
(\ref{flux})
and (\ref{ery}). \\
\noindent
The $\varepsilon$ parameter is defined as:

\begin{equation}
\frac{1}{\varepsilon} = 1 - (4 M^2 + \frac{y}{x} S) {{Q^2+ 4 m_{e}^2}\over
{S^2 (2 - y)^2}} .
\end{equation}
\medskip
For $Q^2 \simeq 0$, $\varepsilon_{R}$ is equal to:
\begin{eqnarray}
\varepsilon_{R} = R \frac{Q^2}{4 m_{e}^2} \simeq 0 
\end{eqnarray}
what simplifies the formula (\ref{anom}) to:
\begin{eqnarray}
{\D^2\sigma_{\rm anom}\over \D y \D Q^2} &=& {\delta_{\rm anom} \sigma^B (y,Q^2)} .
\label{anom1}
\end{eqnarray}
Now it is clear that the contribution from anomalous magnetic moment is a
multiplicative correction to the Born cross section and
changing the Born cross section for DIS to the Born cross section
for $\rho^{0}$ meson production cross section we obtain the $\delta_{\rm anom}$
term in our master formula (\ref{master2}).
 
\medskip\medskip\medskip
\section{ Numerical results for the radiative correction factor for the NMC $\rho^0$ meson production analysis}
In the NMC paper (\cite{nmc}) the results on exclusive $\rho^0$ and $\phi$ 
production in DIS of muons on deuterium, carbon and calcium target is reported.
The experiment was carried out at CERN by NMC collaboration using a 200 GeV muon beam.
To select the $\rho^0$ meson sample the following kinematical cuts 
were used in the analysis (\cite{nmc}): 
\medskip\
$Q^2 \geq 2 GeV^2$ ,
$\nu$ between 40 and 190 GeV and
$y_{max} = 0.9$. 
The cuts for
minimum energy of scattered muon,  
minimum hadron energy ,
minimum $m^2_{e^{+}e^{-}}$
(where  $m^2_{e^{+}e^{-}}$ is the invariant mass for a pair of tracks
calculated assuming the electron mass for both particles) and on the invariant mass of pion's pair ($m_{\pi^+ \pi^-}$)
 were also used in the experimental analysis \cite{nmc}. 
And finally the inelasticity I cut was applied:
\begin{equation}
-0.1 < I < 0.08
\end{equation}
 $I = 0$ is the signal of exclusive $\rho^0$ production.
\medskip
The integration variable $M_h^2$ can be expresed in
terms of the inelasticity I:
\begin{eqnarray}
M_h^2  &=& M^2 + 2 M \nu (1- I) -Q_h^2 . 
\label{nuh}
\end{eqnarray}
The integration limits for $Q_h^2$ are given by the following
relations:
\begin{eqnarray}
Q^2_{h min} &=& Q^2 + 2 \nu I (\nu - \sqrt{\nu^2 + Q^2}) ,\nonumber \\
Q^2_{h max} &=& min\left\{ 2 m \nu, Q^2 + 2 \nu I 
(\nu - \sqrt{\nu^2 + Q^2} \right\}) . 
\label{qh}
\end{eqnarray}

The events with inelasticity I differs from 0 can be considered as the 
radiative events 
or the events with so-called smearing effect \cite{nmc}.
In our aproach we are not able to take into account the smearing effect
and the all events in the specified range of I are treated as a radiative
events. This is not fully correct and therefore our estimates should be rather considered as an upper limit of the correction.\\
The real radiative processes are allowed for $I > 0$. 
Therefore in our calculations the cut used for I was:
\begin{equation}
0 < I < 0.08 
\end{equation}
The most restricted range of inelasticity:
\begin{equation}
-0.05 <  I <  0.0
\end{equation}
was also applied in the NMC analysis for the smaller sample of data; in that case
there is no room for radiative corrections (when smearing is neglected)
from real photons
but still virtual corrections are taken into account.
 
\medskip
The data were parametrized according to: 
\begin{eqnarray} 
\sigma^B_{\gamma^\star \rho^0}= \sigma_0 \left({{Q_0^2}\over{Q^2}}\right)^\beta
\label{param1}
\end{eqnarray}
where $Q_0^2$ was set equal to the average $Q^2$ of the data sample
(for details  see NMC paper \cite{nmc}).
The parameter $\sigma_0$ depends weakly on A of the target (deuterium, carbon or calcium)
in contrast to power
$\beta$ which is practically independent on the target.
Fortunately in our master formula for radiative correction factor $\eta_{\rho^0}$,
eq. (\ref{master1}), the ratio of the cross sections 
(${\sigma^{B}_{\gamma^\star \rho^0}}'$ / $\sigma^B_{\gamma^\star \rho^0}$)
is used and therefore the $\sigma_0$ cancels. It means that the radiative 
corrections are the same for all used targets. 
\medskip
The numerical results for factor $\eta_{\rho^0}$ are presented in Fig.~2
for different $\nu$ and different $Q^2$ bins. The $\beta$ factor and the
$R_{\rho^0}$ are taken from NMC analysis, \cite{nmc} and equal to be
2.02 and 2.0, respectively.
\medskip
The effect of the radiative corrections is varying between 2 and 5 \%.
To estimate more precisely the radiative corrections (e.g. taking into account
the smearing effect)
the Monte Carlo simulations are needed.

\medskip\medskip\medskip
\section{ Numerical results for the radiative correction factor for the ZEUS $\rho^0$ meson photoproduction analysis}

In the ref. \cite{zeus} the measurement of the elastic $\rho^0$
photoproduction cross section at mean $W$ of 70 GeV is reported by ZEUS
collaboration. In the analysis only the momenta of the final state pions were
measured. Events in which the scattered electron was detected in the ZEUS
calorimeter were rejected. It restricts $Q^2$ to the values smaller than $Q^2_{max} \simeq 3-4 GeV^2$.
The minimal value of $Q^2$ is determined by electron mass and the range of
$y$ covered by the measurement:
\begin{equation}
Q^2_{min} = m_e {{y^2}\over{(1-y)}} 
\end{equation}
and is close to $10^{-9}$. The median $Q^2$ is approximately $10^{-4} GeV^2$.
There were several  experimental cuts used in the analysis 
(see \cite{zeus}). In our calculations we used the cut  $60 < W < 80$ GeV 
(which can be simply related to $y$ cut) together with the $Q^2$ limits.
\medskip
The integration limits in the second double integral are defined 
by the set of the following equations:
\begin{eqnarray}
Q_{h min} &=& Q^2 + {{S_{lh} S_{-}}\over{2 M^2}} \nonumber \\
Q_{h max} &=& min\left\{  Q^2 + {{S_{lh} S_{+}} \over{2 M^2}}, S_h \right\} \nonumber \\
S_{lh} &=& S_l - S_h \nonumber \\
S_{-} &=& S_l - \sqrt{S_l^2 + 4 M^2 Q^2} \nonumber \\
S_{+} &=& S_l + \sqrt{S_l^2 + 4 M^2 Q^2} \nonumber \\
M_h^2 &=& M^2 + S_h - Q_h^2 \nonumber \\
S_l &=& S y \nonumber \\ 
S_h &<& S_L
\label{set}
\end{eqnarray}
\medskip
The numerical integrations in eq.(\ref{master2}) are complicated.
The four-dimensional integral with numerically divergent part 
(two infrared divergent parts which should cancel numerically) and 
with the integrand function varying several orders
of magnitude is performed (very small $Q^2$, photoproduction limit).
To simplify the problem part of integrations (over $Q_h^2$) was done 
analytically using {\it Mathematica} system; it allowed to control 
much better 
the divergences cancelation and to decrease computation time.
\medskip
The $\rho^0$ cross section was parametrized as in \cite{zeus} and in 
\cite{DL}. 
The $\eta_{\rho^0}$ radiative correction factor was estimated to be
smaller than 2\%. The nontrivial cancelation between virtual and real
corrections were observed. 

\medskip\medskip\medskip
\section{ Concluding remarks}
In the present paper the method of calculations of the radiative corrections 
for vector mesons exclusive production was presented. The numerical 
calculations were done for $\rho^0$ meson production for two different
experimental analysis: NMC (fixed target 200 GeV muon beam, with virtuality
of photons $Q^2$ higher
than $2 GeV^2$) and
ZEUS at HERA (electron-proton collider, using quasi-real photons with space-like
virtuality $Q^2$ between $10^{-8}$ and $10^{-2} GeV^2 $, at average centre-of-mass energy (W) of 180 GeV).
The large as well as small, close to 0, $Q^2$ ranges are considered.
The method of calculations is based on the approach used in the
calculations of the radiative corrections for DIS (so-called Dubna radiative
correction scheme). 
\medskip
The method can be applied to different experimental analysis
like ZEUS or H1 large $Q^2$ $\rho^0$ production at HERA measurements 
(\cite{h1}).
Our calculations can be also usefull in estimation of radiative correction
effects in $\phi$ meson production or J/$\Psi$ production (e.g. \cite{fipsi}).
The presented method is based on semianalytical approach and
sometime it can be difficult to take into account all experimental cuts
in calculations. In that cases the Monte Carlo approach for radiative 
corrections should be applied.

\medskip\medskip\medskip
\noindent
\newpage
{\Large {\bf {Acknowledgements}}} 

\medskip\medskip\medskip
I thank my colleagues from the NMC and ZEUS for discussions
of radiative corrections and $\rho^0$ meson production. 
I specialy thank Michele Arneodo, Barbara Bade\l ek, Krzysztof ~Muchorowski 
and Andrzej Sandacz. 
This research was supported in part by the Polish Committee for Scientific
Research, grant number  2 P302 069 04.

\noindent
\medskip\medskip\medskip
\section {Appendix A}
\medskip \medskip
Below the exact expressions for certain functions in the D scheme will be given.
\medskip
The 'radiator' functions  ${\cal S}_i$ and the function ${\cal F}^{IR}$ are:
\begin{eqnarray}
{\cal S}_{1}(y,Q^2,y_h,Q^2_h)
&=& \Biggl\{ \frac{1}{\sqrt{C_{2}}} \left[ \frac{Q^2_h-Q^2}{2}
 +\frac{(Q^2+2m^{2})(Q^2_h-2m^{2})}{Q^2_h-Q^2} \right] \nonumber \\
& &-  m^{2}(Q^2_h-2m^{2})\frac{B_{2}}{C_{2}^{3/2}} \Biggr\}
 - \Biggl\{ S \leftrightarrow - X  \Biggr\} +\frac{1}{\sqrt{A_2}}, \nonumber \\
{\cal S}_{2}(y,Q^2,y_h,Q^2_h)
&=& \Biggl\{\frac{1}{\sqrt{C_{2}}} [ M^{2}(Q^2_h + Q^2)-
XS_h] \nonumber \\
& & +
\frac{1}{(Q^2_h-Q^2)\sqrt{C_{2}}} \Biggl[ Q^2_h[S(S-S_h)+X(X+S_h)
 - 2M^{2}(Q^2_h+2m^2) ]  \nonumber \\
& & + 2m^{2} [(S-S_h)(X+S_h)+SX] \Biggr]
\nonumber \\
& &-2m^{2} \frac{B_{2}}{C_{2}^{3/2}} [S(S-S_h)-M^{2}Q^2_h ] \Biggr\}
- \Biggl\{ S \leftrightarrow -X  \Biggr\}
-\frac{2M^{2}}{\sqrt{A_2}},  \nonumber \\
{\cal F}^{IR}(y,Q^2,y_h,Q^2_h) &=&
         \frac{Q^2+2m^{2}}{Q^2-Q^2_h}
         \Bigl( \frac{1}{\sqrt{C_{1}}}-\frac{1}{\sqrt{C_{2}}} \Bigr)\
         -m^{2} \Bigl(\ \frac{B_{1}}{C_{1}^{3/2}}+
         \frac{B_{2}}{C_{2}^{3/2}} \Bigr)\ ,
\nonumber
\end{eqnarray}
where
\begin{eqnarray}
A_2 &=& \lambda_l \equiv A_1,  \nonumber       \\
\lambda_l &=& S_l^2 + 4M^2Q^2  \nonumber \\
B_2 &=& 2 M^2 Q^2 ( Q^2- Q^2_h ) + X(S_lQ^2_h-S_hQ^2)    \nonumber \\
    & & +~SQ^2(S_l-S_h)  \equiv -B_1
    (S \leftrightarrow - X), \nonumber  \\
C_2 &=& [XQ^2_h-Q^2(S-S_h)]^2 + 4m^2
    [(S_l-S_h)(S_lQ^2_h-S_hQ^2) \nonumber \\
     & &-~M^2(Q^2_h - Q^2 )^2]
    \equiv C_1 [S \leftrightarrow -X ],
\nonumber
\end{eqnarray}
with
\begin{eqnarray*}
        X=S(1-y) &=& 2ME' \\
             S_h &=& Sy_h \\
\end{eqnarray*}
and the hadron defined variables, $x_h, y_h$ are given by the following
equations
\begin{eqnarray*}
            M_X^2&=&M^2+Sy_h(1-x_h) \\
            Q^2_h&=&Sx_hy_h  \\
\end{eqnarray*}
\medskip \medskip \medskip
\noindent

\section{Appendix B}
\medskip \medskip
\noindent
Below the exact (with electron's mass terms) expressions for $\delta_{\rm vac}, 
\delta_{R}, \delta^{VR}$ and $\delta_{\rm vtx}$ are given:
         
\begin{eqnarray}
\delta_{\rm vac}^{e,\mu } &=& \frac{2\alpha}{\pi}\left[\frac{-5}{9}+
   \frac{4m_{e,\mu}^2}{3Q^2}+\frac{1}{3}\sqrt{1+\frac{4m_{e,\mu}^2}{Q^2}}
   \left(1-\frac{2m_{e,\mu}^2}{Q^2}\right) \ln \left(
        \frac{\sqrt{1+4m_{e,\mu}^2/Q^2}+1}{\sqrt{1+4m_{e,\mu}^2/Q^2}-1}
           \right) \right]  \nonumber \\ \nonumber \\
\delta_{R} &=& \frac{-\alpha}{\pi} \ln\frac{(W^2 -M^2)^2}{m^2 W^2} \left (
\frac{1+\beta^2}{2 \beta} L_{\rm \beta} -1 \right ) \nonumber \\ \nonumber \\
\delta^{VR} &=& \delta_{\rm vtx} - \frac{\alpha}{\pi} \frac{1+\beta^2} 
{2 \beta} \left [ L_{\rm \beta} \ln{\frac{4 \beta^2}{\beta^2 - 1}} +
\Phi (\frac{1+\beta}{1-\beta}) - \Phi (\frac{1-\beta}{1+\beta}) \right] \nonumber \\
& &+ {\frac{\alpha}{\pi}} {\left [\frac{1}{2 \beta_{1}} 
\ln{\frac{1+\beta_{1}}{1-\beta_{1}}} + \frac{1}{2 \beta_{2}} 
\ln{\frac{1+\beta_{2}}{1-\beta_{2}}} + S_{\rm \Phi} \right ]}    \nonumber 
\\ \nonumber \\
\delta_{\rm vtx} &=& \frac{2 \alpha}{\pi} \left ( -1 + \frac{3}{4} 
\beta L_{\rm \beta} \right ) \nonumber \\ \nonumber \\
S_{\rm \Phi} &=& \frac{1}{2} \left(Q^2 + 2 m^2\right) \int_{0}^{1}\! 
{{d \alpha}\over{
\beta_{\rm \alpha} (-k_{\rm \alpha}^2 )}} \ln{\frac{1-\beta_{\rm \alpha}}{
1+\beta_{\rm \alpha}}}  \nonumber 
\end{eqnarray}

where:
\begin{eqnarray}
\beta &=& \sqrt{1+{{4 m^2} \over {Q^2}}}   \nonumber \\
\beta_{1} &=& \sqrt{1 - {{4 m^2 M^2} \over {(S - Q^2)^2}}} \nonumber \\
\beta_{2} &=& \sqrt{1 - {{4 m^2 M^2} \over {[S (1-y) + Q^2]^2}}} \nonumber \\
L_{\rm \beta} &=& \ln{{\beta +1} \over {\beta -1}} \nonumber
\end{eqnarray}
and finally:
\begin{eqnarray}
\beta_{\rm \alpha} &=& \frac{\mid {\bf k_{\rm \alpha}} \mid}{k_{\rm \alpha}^{0}} \nonumber \\
k_{\rm \alpha} &=& k \alpha + k' (1 - \alpha) \nonumber \\
-k_{\rm \alpha}^2 &=& m^2 + Q^2 \alpha (1 - \alpha) \nonumber
\end{eqnarray}

\medskip \medskip \medskip
\noindent
{\Large {\bf {Figure Captions}}}
\medskip \medskip
\begin{enumerate}
\item Diagram of the exclusive leptoproduction (muon and electron) of the
vector meson.
\item The $\eta_{\rho^0}$ correction factor from eq.(\ref{master1} ) for
different $\nu$ and $Q^2$ bins calculated for NMC (\cite{nmc}).

\end{enumerate}

\end{document}